\documentclass[12pt]{iopart}
\usepackage{graphicx}
\begin{document}

\hfill  DAMTP-2015-44

\title{Twistors and the  massive spinning particle}

\author{Luca Mezincescu}
\address{Department of Physics, University of Miami,\\
Coral Gables, FL 33124, USA}
\ead{mezincescu@server.physics.miami.edu}

\author{Alasdair J. Routh and Paul K. Townsend}
\address{Department of Applied Mathematics and Theoretical Physics,\\ Centre for Mathematical Sciences, University of Cambridge,\\
Wilberforce Road, Cambridge, CB3 0WA, U.K.}
\ead{A.J.Routh@damtp.cam.ac.uk, P.K.Townsend@damtp.cam.ac.uk}
\vspace{10pt}
\begin{abstract}
Gauge-invariant twistor variables are found for the massive spinning particle with ${\cal N}$-extended  local worldline supersymmetry, in spacetime dimensions $D=3,4,6$. The twistor action is manifestly Lorentz invariant  but  the anticommuting spin variables appear exactly as  in the non-relativistic limit.  This allows a simple confirmation that the quantum ${\cal N}=2$ spinning particle has either spin one or spin zero, and that  ${\cal N}>2$ is quantum inconsistent  for $D=4,6$. 
\end{abstract}

%
%
%
%
%

\font\mybb=msbm10 at 12pt
\def\bb#1{\hbox{\mybb#1}}
\def\bZ {\bb{Z}}
\def\bR {\bb{R}}
\def\bE {\bb{E}}
\def\bT {\bb{T}}
\def\bM {\bb{M}}
\def\bL {\bb{L}}
\def\bG {\bb{G}}
\def\bH {\bb{H}}
\def\bC {\bb{C}}
\def\bA {\bb{A}}
\def\bK {\bb{K}}
\def\bO {\bb{O}}
\def\bX {\bb{X}}
\def\bP {\bb{P}}
\def\bJ {\bb{J}}
\def\bU {\bb{U}}
\def\bJ {\bb{J}}
\def\bV{\bb{V}}
\def\bW {\bb{W}}
\def\bI {\bb{I}}

\def\bfo{\mbox{\boldmath $\omega$}}
\def\bfO{\mbox{\boldmath $\Omega$}}
\def\bfn{\mbox{\boldmath $\nabla$}}
\def\bfs{\mbox{\boldmath $\sigma$}}
\def\bft{\mbox{\boldmath $\tau$}}
\def\bfS{\mbox{\boldmath $\Sigma$}}
\def\bfpsi{\mbox{\boldmath $\psi$}}
\def\bfG{\mbox{\boldmath $\Gamma$}}
\def\bfth{\mbox{\boldmath $\theta$}}
\def\bfTh{\mbox{\boldmath $\Theta$}}
\def\bfPi{\mbox{\boldmath $\Pi$}}
\def\bfpi{\mbox{\boldmath $\pi$}}
\def\tr{{\rm tr}}


\newcommand{\sect}[1]{\setcounter{equation}{0}\section{#1}}
\renewcommand{\theequation}{\arabic{section}.\arabic{equation}}

\section{Introduction}
\setcounter{equation}{0}

The spacetime dimensions $D=3,4,6$ have the special property that there is an extension of the conformal group, available for all $D$, to a superconformal group \cite{Nahm:1977tg}.  This is relevant to  massless supersymmetric field theories, and free field theories  of this type arise from quantization of the massless superparticle \cite{Brink:1981nb,Siegel:1983hh}, which is superconformal invariant precisely in dimensions $D=3,4,6$. The superconformal invariance of the massless superparticle action can be made manifest by a formulation \cite{Shirafuji:1983zd,Bengtsson:1987si}
 in  which the phase space  is parametrized by a supertwistor \cite{Ferber:1977qx}. Spin-shell constraints then replace the usual mass-shell constraint, ensuring that  the physical phase space dimension is unchanged,  and also that the superparticle describes (upon quantization) a supermultiplet of zero super-helicity. 

Although the action for a {\it massive} particle cannot be conformal invariant, the massive superparticle \cite{Casalbuoni:1976tz} still has a  supertwistor formulation in dimensions $D=3,4,6$  (which we abbreviate to 3D etc.) but now the phase superspace is parametrized by {\it two} supertwistors. The necessity of doubling the twistor phase space was initially discovered  in the context of the  twistor approach to solutions  of massive wave equations \cite{Hughston:1981zc}. One expects to recover such results by covariant quantization of the twistor formulation of corresponding massive particle mechanics models, which is our focus here.

The twistor formulation of at least some massive particle mechanics models may be found indirectly by ``dimensional reduction'' of known twistor formulations of massless particle mechanics models  in a higher dimension; in this context, ``dimensional reduction'' amounts to the incorporation of a constraint on the particle's momentum in the extra dimensions. This naturally leads to a  doubled twistor phase space in the lower dimension because (i) a twistor is a spinor of the conformal group \cite{Penrose:1986ca}, which decomposes into a pair of spinors of the lower-dimensional conformal group, and (ii) the conformal invariance in the lower dimension is broken  only by the additional momentum constraint,  which has no influence on the nature of the phase space.  

This construction was first used in \cite{deAzcarraga:2008ik}: the supertwistor formulation of the massive 4D superparticle was found  by reduction of the known supertwistor formulation of the massless 6D superparticle.  In an earlier article, we reviewed and extended  the known  results on this topic  \cite{Mezincescu:2013nta},  and two of us have recently found (by a direct method)  a supertwistor formulation of the massive 6D superparticle \cite{Routh:2015ifa}. As explained in that work, the combined results for the $D=3,4,6$ massive superparticle  fit  nicely with the idea  \cite{Kugo:1982bn,Sudbery:1983}  that properties of supersymmetric theories in spacetime dimensions $D=3,4,6,10$ are related to the division algebras $\bR,\bC,\bH,\bO$. 

 It has been known for a long time \cite{Townsend:1991sj} that there is also a supertwistor formulation of the 4D massless ``spinning particle'' \cite{Brink:1976sz,Brink:1976uf}, which has  local worldline supersymmetry rather than global spacetime supersymmetry.  The spinning particle  action in supertwistor  variables is  remarkably similar to that of the superparticle  but the spin-shell constraints are slightly different, breaking superconformal invariance  to conformal invariance and leading to a quantum theory with states of  a spin-1/2 particle  rather than a spacetime supermultiplet.  This  result was generalized in \cite{Mezincescu:2013nta} to the ${\cal N}$-extended massless 4D spinning particle \cite{Gershun:1979fb,Howe:1988ft}, which describes a particle of spin  $\frac{1}{2}{\cal N}$, and the results were then used to find analogous results for the {\it massive} 3D spinning particle.  However, the constructions underlying these results do not appear to apply more generally.
 
 In this paper we present a  twistor formulation of the massive spinning particle in $D=3,4,6$, for any ${\cal N}$.  Our 3D  results  duplicate those of \cite{Mezincescu:2013nta} but our improved construction generalizes to both 4D and 6D.  We say ``twistor'' rather than supertwistor because the anticommuting phase-space variables  turn out to be different (for $D=4,6$) from those of  the superparticle.  Implicit in our results is a twistor formulation of the {\it massless} spinning particle for $D=3,4,6$. We present the details for $D=3$, showing how conformal invariance is recovered in the massless limit. For $D=4$ 
the analogous final result differs slightly from   \cite{Townsend:1991sj}  because the starting point there was the standard form of the massless spinning particle, which differs from what one gets by taking the zero-mass limit of the standard massive spinning particle action. 

The main point of  our twistor reformulation of  massive ``spinning particle''  mechanics is that  {\it the twistor variables are gauge invariant with respect to local worldline supersymmetry}. New gauge invariances are  introduced, but not ``fermionic'' ones, which means that  all twistorial anticommuting variables are physical; they appear in the action in exactly the same way that they would in the analogous {\it non-relativistic} action!  This feature simplifies the determination of some properties of the quantum theory, in particular for the ${\cal N}$-extended spinning particle.

For ${\cal N}>2$ it is known (and we confirm)  that the massive spinning particle model is  inconsistent for even $D$ because of a global anomaly. This follows, as pointed out in  \cite{Howe:1989vn}, from a global anomaly exhibited there for the massless ${\cal N}>2$ superparticle in odd $D$. This problem can be evaded for ${\cal N}=2$  because in this case it is possible to cancel the anomaly by adding a worldline Chern-Simons (WCS) term with half-integral coefficient \cite{Howe:1989vn}. We use our results to confirm that  the ${\cal N}=2$ massive 4D superparticle describes either a spin-zero or a spin-one particle depending on the choice of WCS coefficient. 

We should mention here  that our initial motivation for considering twistor formulations of massive particle mechanics models was a similarity to twistor formulations of the Nambu-Goto 
string \cite{Shaw:1986dq,Cederwall:1989su}. In both cases a doubling of the twistor phase space is needed (compared to a massless particle). The constructions described here for the 
massive spinning particle may therefore be useful in any future attempt to find a twistor formulation of the spinning string. 

We begin with a summary of the ${\cal N}$-extended spinning particle action in arbitrary spacetime dimension $D$. We then proceed to its twistor formulation 
for $D=3,4,6$, dealing sequentially with these dimensions.  For 3D we discuss only the ${\cal N}=1$ case; this suffices to introduce the new construction
and some generic features of our spinor conventions.  For 4D we first discuss the  ${\cal N}=1$ case and then  generalize to ${\cal N}>1$, using the results to discuss the quantum theory. For 6D we take over some results of  \cite{Routh:2015ifa} for the bosonic particle, making explicit some conventions implicit in that work, and then present  the twistor form of the massive spinning particle.  We conclude with a discussion of some general features of our results.    

Finally, we include an appendix in which the supertwistor form of the 4D superparticle action is found in the conventions of this paper. This is essentially a more elegant version of previous results but we also keep track of the sign of the energy in the solution to the mass-shell constraint in order to illustrate an important difference between spinning particles and superparticles.

\section{Spinning particle preliminaries}
\setcounter{equation}{0}

For any spacetime dimension, the phase-space action for the minimal  massive spinning particle, with ${\cal N}=1$ worldline supersymmetry, is  
\begin{equation}
\!\!\!\!\!\!\!\!\!\!\!\!\!\!\!\! S= \int \! dt\left\{ \dot X^m P_m + \frac{i}{2}\lambda^m\dot\lambda_m + \frac{i}{2}\xi\dot \xi - \frac{1}{2} e\left(P^2+m^2\right) + i\zeta \left(\lambda^m P_m +m\xi\right)\right\}\, . 
\end{equation}
We use here, and throughout the paper, the Minkowski metric with ``mostly plus''  signature. 
The canonical variables $(\lambda^m,\xi)$ are anticommuting, as is the Lagrange multiplier $\zeta$. The Hamiltonian constraints are both first class, and they generate the gauge transformations
\begin{equation}
\delta X^m = a P^m -i\epsilon\lambda^m \, , \qquad 
\delta \lambda^m = P^m \epsilon \, , \qquad
\delta\xi= m\epsilon\, , 
\end{equation}
for commuting parameter $a(t)$ and anticommuting parameter $\epsilon(t)$. The action is invariant if the Lagrange multipliers are assigned the transformations
\begin{equation}
\delta e= \dot a -2i\epsilon\zeta\, , \qquad \delta\zeta = \dot\epsilon\, . 
\end{equation}
This model describes, upon quantization, a massive spin-1/2 particle \cite{Brink:1976uf},  the fermionic constraint becoming the Dirac equation when imposed as a physical state condition. 
Notice that the bosonic phase space is spanned by two $D$-vectors subject to one first-class constraint, to which we must add one anticommuting $D$-vector and one anticommuting scalar that are also subject to one first-class constraint. This leads to a physical phase superspace with graded dimension $(2D-2|D-1)$. 

If one takes the $m=0$ limit of the above results then one arrives at the action 
\begin{equation}\label{masslessact}
\!\!\!\!\!\!\! S= \int \! dt\left\{ \dot X^m P_m + \frac{i}{2}\lambda^m\dot\lambda_m + \frac{i}{2}\xi\dot \xi - \frac{1}{2} e\, P^2 + i\zeta \, \lambda^m P_m \right\}\, . 
\end{equation}
We could omit the $\xi\dot\xi$ term because $\xi$ is now inert under the local worldline supersymmetry, and it has dropped  out of the constraints. Indeed, this term {\it is} omitted from the standard ``massless spinning particle'' action, but omitting it introduces a discontinuity into the massless limit of the massive spinning particle. 

The ${\cal N}$-extended massive spinning particle is constructed by incorporating more anticommuting variables. Specifically, we make the replacement
\begin{equation}
\lambda^m \to \lambda^m_a\, , \quad \xi \to \xi_a\, , \quad \zeta\to \zeta_a \, \qquad \left(a=1,\dots,{\cal N}\right)\, , 
\end{equation}
and then gauge the resulting $SO({\cal N})$ symmetry by  including additional Hamiltonian constraints, imposed by means of  a  new commuting antisymmetric $SO({\cal N})$-tensor Lagrange multiplier $f_{ab}$. The resulting action is  \cite{Gershun:1979fb,Howe:1988ft}
\begin{eqnarray}
S &=& \int \! dt\Big\{ \dot X^m P_m + \frac{i}{2}\lambda_a \cdot \dot\lambda_a + \frac{i}{2}\xi_a\dot \xi_a - \frac{1}{2} e\left(P^2+m^2\right) +  i\zeta_a \left(\lambda_a\cdot P +m\xi_a\right) \nonumber \\
&& \qquad \qquad \qquad -\, i f_{ab} \left(\lambda_a \cdot \lambda_b + \xi_a\xi_b\right)\Big\}\, ,  
\end{eqnarray}
where summation over the $SO({\cal N})$ indices is implicit. This  action has a local $SO({\cal N})$ gauge invariance in addition to its ${\cal N}$ local 
worldline supersymmetries. 

The ${\cal N}=2$ case is special because then $f_{ab}\propto \epsilon_{ab}f$ with an  $SO(2)$ transformation of $f$ that is a total derivative, 
allowing us to add to the action a term linear in $f$; this is the worldline Chern-Simons  term. We now have the action 
\begin{eqnarray}
S &=& \int \! dt \Big\{ \dot X^m P_m + \frac{i}{2}\lambda_a \cdot \dot\lambda_a + \frac{i}{2}\xi_a\dot \xi_a - \frac{1}{2} e\left(P^2+m^2\right) + i\zeta_a \left(\lambda_a\cdot P + m\xi_a\right) \nonumber \\
&& \qquad\qquad\qquad - f\left[ \epsilon^{ab} \left(\lambda_a \cdot \lambda_b + \xi_a\xi_b\right) -c\right] \Big\}\, , 
\end{eqnarray}
where the constant $c$ is the coefficient of the WCS term; it  is quantized in the quantum theory  with choices that lead  to a particle mechanics description of massive $p$-form fields for some integer $p$ \cite{Howe:1989vn}.

\section{Massive 3D spinning particle}
\setcounter{equation}{0}

We shall use a notation in which spacetime vectors are bi-spinors of $Sl(2;\bR)$.  We spell out our conventions here because some important features
that carry over to  4D and 6D are simpler to discuss for 3D.  We start from the real  $2\times2$ Dirac matrices\footnote{This differs slightly from the choice made in \cite{Mezincescu:2013nta}.}
\begin{equation}\label{gamma}
\left(\gamma^0\right)^\alpha{}_\beta  = i\sigma_2 \, , \qquad \left(\gamma^1\right)^\alpha{}_\beta =  \sigma_3\, , \qquad \left(\gamma^2\right)^\alpha{}_\beta =-\sigma_1\, .  
\end{equation}
These matrices satisfy the identities
\begin{equation}
\left(\gamma^m\right)^\alpha{}_\beta \left(\gamma_m\right)^\gamma{}_\delta \equiv 2\delta^\alpha_\delta\delta^\gamma_\beta - \delta^\alpha_\beta\delta^\gamma_\delta \, , \qquad \left(\gamma^m\right)^\alpha{}_\beta\left(\gamma^n\right)^\beta{}_\alpha \equiv 2\eta^{mn}\, , 
\end{equation}
where $\eta$ is the (mostly plus) Minkowski metric. Spinor indices will  be raised or lowered by means of  the alternating invariant tensor $\varepsilon$ of $Sl(2;\bR)$, 
using the conventions 
\begin{equation}
\psi^\alpha = \varepsilon^{\alpha\beta} \psi_\beta \, , \qquad \psi^\beta\varepsilon_{\beta\alpha} = \psi_\alpha\, , 
\end{equation}
for any spinor $\psi$.  We define $\varepsilon$ such that 
\begin{equation}
\varepsilon^{\alpha\beta}\varepsilon_{\alpha\gamma}= \delta_\gamma^\beta \qquad \left(\varepsilon^{12}=\varepsilon_{12} =1\right). 
\end{equation}
As observed by Howe  in the context of $Sl(2;\bC)$ spinor index conventions  \cite{Howe:1981gz}, these definitions have the advantage  that  $\varepsilon_{\alpha\beta}$ and  $\varepsilon^{\alpha\beta}$ are related by raising/lowering both indices, so that $\varepsilon$ can be consistently interpreted as an $Sl(2;R)$ tensor. 

Lowering the indices of the Dirac matrices amounts to a multiplication from the left by $-i\sigma_2$, so that 
\begin{equation}
\gamma^m_{\alpha\beta} = \left(1,\sigma_1,\sigma_3\right)\, . 
\end{equation}
Notice that these are symmetric.  Now, for  any Lorentz 3-vector $V$, we define 
\begin{equation}
V_{\alpha\beta} = \gamma^m_{\alpha\beta} V_m  \quad \Leftrightarrow \quad
V^m = -\frac{1}{2}\gamma^m_{\alpha\beta}V^{\alpha\beta}  \, . 
\end{equation}
It follows that 
\begin{equation}
V^2 = - \frac{1}{2} V^{\alpha\beta}V_{\alpha\beta} = - \det V_{\alpha\beta}\, . 
\end{equation}

Using these conventions, we find that the ${\cal N}=1$ 3D massive spinning particle action  is 
\begin{eqnarray}\label{act3}
S&=& \int \! dt \left\{- \frac{1}{2} \dot X^{\alpha\beta}P_{\alpha\beta} - \frac{i}{4} \lambda^{\alpha\beta}\dot\lambda_{\alpha\beta} + \frac{i}{2}\xi\dot\xi 
- \frac{1}{2}e\left(P^2+m^2\right) \right. \nonumber \\
&& \qquad \qquad \qquad \qquad \left.
- \, \frac{i}{2}\zeta \left(P^{\alpha\beta}\lambda_{\alpha\beta} -2m \xi\right)\right\}\, . 
\end{eqnarray}
The  Poisson brackets that follow from this action are 
\begin{equation}
\!\!\!\!\!\!\!\!\!\!\!\! \left\{X^{\gamma\delta}, P_{\alpha\beta}\right\}_{PB}= -2\delta^{(\gamma}_\alpha\delta^{\delta)}_\beta \, , \quad 
\left\{\lambda^{\gamma\delta}, \lambda_{\alpha\beta}\right\}_{PB} = 2i \delta^{(\gamma}_\alpha\delta^{\delta)}_\beta \, , 
\quad \left\{\xi,\xi\right\}_{PB}=-i\, . 
\end{equation}
 It should be remembered that the Poisson bracket of two anticommuting variables is {\it symmetric}  under their interchange.
Using these Poisson bracket relations,  it may be verified that the constraints are first-class and that they generate the gauge  transformations 
\begin{equation}
\delta X^{\alpha\beta} = aP^{\alpha\beta} -i\epsilon\lambda^{\alpha\beta} \, ,\qquad \delta\lambda_{\alpha\beta} = P_{\alpha\beta}\epsilon\, , 
\qquad \delta\xi=m\epsilon\, . 
\end{equation}

In addition to its gauge invariances, the action is manifestly Poincar\'e invariant. The  Noether charges  for translations and Lorentz rotations are, respectively, 
\begin{equation}
{\cal P}_{\alpha\beta} = P_{\alpha\beta}\, , \qquad 
{\cal J}_\alpha{}^\beta = X^{\beta\gamma}P_{\alpha\gamma} - \frac{1}{2} \delta_\alpha^\beta X^{\gamma\delta}P_{\gamma\delta}
-\frac{i}{2} \lambda^{\beta\gamma}\lambda_{\alpha\gamma}\, . 
\end{equation}
One may construct from these Poincar\'e charges the Pauli-Lubanski pseudoscalar; in spinor notation this is 
\begin{equation}
{\cal W} = \frac{1}{2}{\cal P}^{\alpha\beta}{\cal J}_{\alpha\beta}\, . 
\end{equation}

To pass to the twistor form of the action we first solve the mass-shell constraint  by expressing the momentum in terms of a pair of commuting spinors
$U_\alpha^I$ ($I=1,2)$:
\begin{equation}
P_{\alpha\beta} = \mp U_\alpha^IU_\beta^J\delta_{IJ}\, , \qquad \det U =m \, . 
\end{equation} 
The choice of  top sign leads to $P^0>0$, i.e. positive energy; we allow for either sign of the energy since, ultimately,  both positive and negative energies are needed in relativistic field theory.  The constraint on the determinant of the $2\times 2$ matrix with entries $U_\alpha{}^I$ is essentially another version of the mass-shell constraint since $P^2=-\det P = -( \det U)^2$.

Next, we solve the fermionic constraint by writing $\lambda$ in terms of $\xi$ and two new anticommuting variables in the form of a symmetric and traceless 
$SO(2)$ tensor $\psi_{IJ}$:
\begin{equation}
\lambda_{\alpha\beta} = \frac{1}{m}U_\alpha^I U_\beta^J \psi_{IJ} + 
\frac{1}{m}P_{\alpha\beta}\, \xi \, , \qquad \psi_{IJ}=\psi_{JI}\, , \quad \delta^{IJ} \psi_{IJ}=0\, . 
\end{equation}
Notice that the supersymmetry transformation of $\xi$ now implies that of $\lambda$, so that the new anticommuting variables $\psi_{IJ}$ are 
(like $U$) inert under the local worldline supersymmetry.  

Substitution for $P$ and $\lambda$ yields the Lagrangian
\begin{equation}\label{Lag}
L= \dot U_\alpha ^I W^\alpha_I + \frac{i}{4} \psi^{IJ}\dot\psi_{IJ} + \frac{d}{dt}\left(\dots\right)\, , 
\end{equation}
where  
\begin{equation}\label{Wexpress}
W^{\alpha\, I} = \mp X^{\alpha\beta} U_\beta^I  \mp \frac{i}{m} U^\alpha_K 
 \left[ \varepsilon^{JK}\psi_{IJ}\xi \mp \frac{1}{2}\varepsilon^{JL}\psi_{KL}\psi_{IJ}\right]  \, . 
 \end{equation}
 From this expression we may derive the identity
\begin{equation}
0 \equiv \Lambda := \epsilon_{IJ} \left[U_\alpha ^I W^{\alpha\, J} + \frac{i}{2} \psi_{IK}\psi_{KJ}\right]\, . 
\end{equation}
It should be appreciated here that (in contrast to the 4D case to follow) there is no significance to the position (up or down) of the  $I,J,K,L$  indices since they are raised or lowered using the Kronecker delta. 

We see from (\ref{Lag}) that $W$ is canonically conjugate to $U$ but, as things stand, it is not an independent variable. In order to be able to consider it as an independent variable  we must impose $\Lambda=0$ as a constraint by means of a Lagrange multiplier. We must also impose the new mass-shell constraint 
\begin{equation}
0= \varphi := \det U -m \, . 
\end{equation}
In this way we arrive at the action
\begin{equation}\label{3Dspinact}
S= \int\! dt \left\{ \dot U_\alpha ^I W^\alpha_I + \frac{i}{4} \psi^{IJ}\dot\psi_{IJ}  - s\Lambda - \rho\varphi\right\}\, , 
\end{equation}
where $s$ and $\rho$ are the Lagrange multipliers.  This action is manifestly $SO(2)$ invariant, and this is a gauge invariance because of the 
constraint $\Lambda=0$. The new mass-shell constraint  function $\varphi$  is the generator for  the gauge transformations
\begin{equation}\label{atrans}
\delta W^{\alpha\, I} = b\, \epsilon^{IJ} U^\alpha_J\, , \qquad \delta\rho= \dot b\, , 
\end{equation}
for parameter $b(t)$.  This transformation is equivalent to a time reparametrization. We do not present the proof, which involves consideration of  ``trivial'' gauge invariances, because it is very similar to the proof  in  \cite{Routh:2015ifa} for an analogous gauge invariance of the  twistorial 6D superparticle. 

In the absence of the constraints,  the action (\ref{3Dspinact}) would be invariant under the infinite-dimensional group of canonical transformations of the coordinates of a phase superspace of graded dimension $(8|3)$, but this is broken by the $\Lambda=0$ constraint to $Sp(4;\bR)\times SO(2)$. Each of the two spinor pairs $Z^I=(U^I,W_I)$ forms an irreducible 4-plet of $Sp(4;\bR)$ and together they form a doublet of $SO(2)$. As $Sp(4;\bR)$ is a cover of the 3D conformal group such that the $4$-plet of $Sp(4;\bR)$ is its spinor representation, this makes $Z^I$ a pair of 3D twistors, so the bosonic phase space  is parametrized by a pair of 3D twistors, as claimed. However,  the mass-shell constraint  $\varphi=0$ breaks  $Sp(4;\bR)$ to  $Sl(2;\bR)$, i.e. to the Lorentz group,  as expected for a massive particle. 

What we wish to emphasize about the action (\ref{3Dspinact}) is that {\it there is no trace of the local worldline supersymmetry of the action (\ref{act3}) from which we started}. The new variables $U$ and 
$\psi$ are manifestly inert under the local supersymmetry but  it is not obvious that $W$ is too. Initially, $W$ was given by the expression (\ref{Wexpress}); from this expression we may compute the local supersymmetry transformation of $W$ from the known transformations of $X$ and $\xi$. Using the identity
\begin{equation}
\varepsilon^{IK}\psi_{KJ} \equiv \varepsilon^{JK}\psi_{KI}\, , 
\end{equation}
one finds that 
\begin{equation}
\delta_\epsilon W^{\alpha\, I} =-i\epsilon\xi \, \varepsilon^{IJ} U^\alpha_J\, . 
\end{equation}
Although this is non-zero, it has the form of (\ref{atrans}) with a parameter $b= -i\epsilon\xi$. Thus, the new variables $(U,V,\psi)$ are gauge invariant with respect to the original gauge transformations modulo a gauge transformation associated to the new mass-shell constraint. 

The Pauli-Lubanski pseudoscalar in twistor variables is, after using the mass-shell and spin-shell constraints, 
\begin{equation}
{\cal W}= \pm m \Sigma\, , \qquad \Sigma = \frac{i}{4} \epsilon^{IJ}\psi_{IK}\psi_{JK}\, . 
\end{equation}
The two independent components of $\psi_{IJ}$ can be traded  for a single {\it complex}  anticommuting variable $\psi$, such that 
$\{\psi,\bar\psi\}_{PB}=-i$, by writing
\begin{equation}\label{psis}
\psi_{11}=-\psi_{22} = \frac{1}{\sqrt{2}}\left(\psi+\bar\psi\right) \, , \qquad \psi_{12}=\psi_{21} = \frac{1}{\sqrt{2}\, i}\left(\psi - \bar\psi\right)\, . 
\end{equation}
On passing to the quantum theory, $\psi\to \hat\psi$ and $\bar\psi\to \hat\psi^\dagger$, we have  $\{\hat\psi,\hat\psi^\dagger\}=1$ (in units for which $\hbar=1$) 
and hence
\begin{equation}
\hat\Sigma = \hat n- \frac{1}{2}\, , \qquad \hat n = \hat\psi^\dagger\hat\psi\, . 
\end{equation}
The operator  $\hat n$ is a fermi number operator with eigenvalues $0,1$.  The eigenvalues of the spin operator $\hat\Sigma$ are therefore $\pm 1/2$.
We thus confirm, for $3D$,  that the ${\cal N}=1$ massive spinning particle action describes a particle of spin $1/2$. The 
${\cal N}>1$ case was dealt with  in \cite{Mezincescu:2013nta}.

\subsection{Massless limit}

For the 3D case we shall present an analysis of the $m=0$ limit. Setting $m=0$  in  (\ref{3Dspinact}) changes only the mass-shell constraint, which is now $\det U=0$. 
This constraint implies that  the spinors  $\{U^I; I=1,2\}$ are linearly dependent. Assuming non-zero $U^1$, for simplicity of presentation, we then have 
\begin{equation}\label{massless1}
U^1 = U\, , \qquad U^2 = \lambda U\, , 
\end{equation}
for spinor $U(t)$ and scalar $\lambda(t)$.  We may now solve the $\Lambda=0$ constraint by setting
\begin{equation}\label{massless2}
W^1 = W\, , \qquad W^2 = \lambda W + \kappa U  + V\, , 
\end{equation}
for spinor $W(t)$, another scalar $\kappa(t)$, and any spinor $V(t)$ such that 
\begin{equation}\label{Vpsis}
U_\alpha V^\alpha = -2\bar\psi\psi\, . 
\end{equation}
A solution for $V$ exists  because we are assuming non-zero $U$. We may add to $V$ any multiple of $U$ but the solution for $V$ is unique  if we consider it to 
represent the equivalence class  for which $V$ and $V'$ are identified if they differ by a multiple of $U$. 

Gauge invariance of the relation (\ref{Vpsis}) requires the following gauge transformation for $V$
\begin{equation}
\delta V= - \alpha \lambda V\, ,
\end{equation}
where $\alpha(t)$ is the  scalar parameter for $SO(2)$ gauge transformations. Because of the equivalence relation on $V$, it is inert under the 
transformation (\ref{atrans}) with parameter $b(t)$.  Gauge invariance of the relations  (\ref{massless1}) and  (\ref{massless2}) requires the scalars 
$(\lambda,\kappa)$ to transform as follows:
\begin{equation}
\delta \lambda = -(1+\lambda^2)\alpha\, , \qquad \delta\kappa = -2\alpha\lambda \kappa -b(1+\lambda^2)\, . 
\end{equation}
This shows that we may fix the gauge invariances by setting $\lambda=\kappa=0$. At this point the only non-zero independent variables are  $(U,W)$ and $\psi$,  and 
the action reduces to 
\begin{equation}
S= \int dt\left\{ \dot U_\alpha W^\alpha + i \bar\psi\dot\psi\right\}\, . 
\end{equation}
This is the twistor action for the massless 3D spinning particle. It is  manifestly $Sp(4;\bR)$ invariant, with $(U,W)$ transforming as a 4-plet, i.e. as a 3D twistor. 
Notice that the graded dimension of the phase space is $(4|2)$, which is what we should expect from a comparison with the action (\ref{masslessact}).

\section{Massive 4D spinning particle}
\setcounter{equation}{0}

We shall use a notation in which spacetime vectors are bi-spinors of $Sl(2;\bC)$.  Specifically, for any Lorentz 4-vector V, 
\begin{equation}
V^m = -\frac{1}{2}\sigma^m_{\alpha\dot\alpha}V^{\alpha\dot\alpha} \, , \qquad  V^{\alpha\dot\alpha} = V^m\sigma_m^{\alpha\dot\alpha} \, . 
\end{equation}
Here, $\sigma^m =(\bI_2, \bfs)$ where $\bfs$ is the triplet of $2\times2$ Pauli matrices, and 
\begin{equation}\label{sigup}
\sigma_n^{\alpha\dot\alpha} := \eta_{nm}\,  \varepsilon^{\alpha\beta}\varepsilon^{\dot\alpha\dot\beta} \sigma^m_{\beta\dot\beta}\, . 
\end{equation}
The $\sigma$ matrices satisfy the relations 
\begin{equation}
\eta_{mn} \, \sigma^m_{\alpha\dot\alpha} \sigma^n_{\beta\dot\beta} = -2\varepsilon_{\alpha\beta}\varepsilon_{\dot\alpha\dot\beta}\, , \qquad 
\sigma^m_{\alpha\dot\alpha} \sigma_n^{\alpha\dot\alpha} = -2\delta^m_n\, , 
\end{equation}
where $\eta$ is the Minkowski metric, which we take to have ``mostly plus'' signature.

As (\ref{sigup}) suggests,  spinor indices are raised and lowered by means of the  $Sl(2;\bC)$ invariant alternating tensors. As in 3D, we do this according to the convention that, for any spinor $\psi$, 
\begin{equation}
\varepsilon^{\alpha\beta}\psi_\beta = \psi^\alpha\, , \qquad  \varepsilon^{\dot\alpha\dot\beta}\psi_{\dot\beta} = \psi^{\dot\alpha}\, \qquad
\psi^\alpha\varepsilon_{\alpha\beta} = \psi_\beta\, , \qquad \psi^{\dot\alpha} \varepsilon_{\dot\alpha\dot\beta} = \psi_{\dot\beta}\, .
\end{equation}
For any Lorentz vector $V$ we have
\begin{equation}
\eta^{mn}V_m V_n \equiv V^2 = - \frac{1}{2} V^{\alpha\dot\beta}V_{\alpha\dot\beta} = -\det V\, , 
\end{equation}
where the last equality follows from the definition of the determinant of the $2\times 2$ matrix with entries $V_{\alpha\dot\alpha}$. The factors here are  a reflection of the fact that $\eta^{\alpha\dot\alpha, \beta\dot\beta} V_{\beta\dot\beta} = -2 V^{\alpha\dot\alpha}$.

In these spinor conventions,  the action for the ${\cal N}=1$ spinning particle is 
\begin{eqnarray}
S &=& \int\! dt \left\{-\frac{1}{2} \dot X^{\alpha\dot\alpha} P_{\alpha\dot\alpha} - \frac{i}{4} \lambda^{\alpha\dot\alpha} \dot\lambda_{\alpha\dot\alpha} + \frac{i}{2} \xi\dot\xi - \frac{1}{2} e\left(P^2 +m^2 \right) \right.\nonumber \\
&& \qquad\qquad \qquad \qquad \qquad \left.
- \, \frac{i}{2} \zeta\left(P^{\alpha\dot\alpha} \lambda_{\alpha\dot\alpha} - 2m \xi\right) \right\}\, . 
\end{eqnarray}
The canonical Poisson brackets are
\begin{equation}
\!\!\!\!\!\! \!\!\!\!\!\!   \left\{ X^{\beta\dot\beta},P_{\alpha\dot\alpha}\right\}_{PB} = -2\delta_\alpha^\beta\delta_{\dot\alpha}^{\dot\beta}\, , \quad 
\left\{\lambda^{\beta\dot\beta},\lambda_{\alpha\dot\alpha}\right\}_{PB} = 2i \delta_\alpha^\beta\delta_{\dot\alpha}^{\dot\beta}\, , 
\quad \left\{\xi,\xi\right\}_{PB} =-i\, . 
\end{equation}
The gauge transformations of the canonical variables are now 
\begin{equation}\label{deltas1}
\delta X^{\alpha\dot\alpha} = aP^{\alpha\dot\alpha} -i\epsilon \lambda^{\alpha\dot\alpha}\, , \qquad \delta\lambda_{\alpha\dot\alpha}= P_{\alpha\dot\alpha}\epsilon\, , \qquad \delta\xi =m\epsilon\, . 
\end{equation}

In addition to its gauge invariances, the action is also invariant under the Poincar\'e isometries of 4D  Minkowski space. The corresponding Noether charges are ${\cal P}_{\alpha\dot\alpha} = P_{\alpha\dot\alpha}$ for  translations, and 
\begin{eqnarray}
{\cal J}_\alpha{}^\beta &=& \frac{1}{2} X^{\beta\dot\alpha} P_{\alpha\dot\alpha} - 
\frac{1}{4} \delta_\alpha^\beta X^{\gamma\dot\gamma}P_{\gamma\dot\gamma}  - 
\frac{i}{4}\lambda^{\beta\dot\alpha}\lambda_{\alpha\dot\alpha}\, , \nonumber \\
\bar{\cal J}_{\dot\alpha}{}^{\dot\beta} &=& \frac{1}{2} X^{\alpha\dot\beta}P_{\alpha\dot\alpha} -
 \frac{1}{4}\delta_{\dot\alpha}^{\dot\beta} X^{\gamma\dot\gamma}P_{\gamma\dot\gamma}  - 
 \frac{i}{4} \lambda^{\alpha\dot\beta}\lambda_{\alpha\dot\alpha}
\end{eqnarray}
for Lorentz rotations.

To construct the twistor form of the action, we proceed as in the 3D case. We first solve the mass-shell constraint by writing $P$
 in terms of the $U(2)$ doublet of commuting complex spinors $U_\alpha^I$ ($I=1,2$), and their complex conjugates  $\bar U_{\dot\alpha\, I}$:
\begin{equation}\label{Pus}
P_{\alpha\dot\alpha} = \mp U_{\alpha}^I\bar U_{\dot\alpha\, I}\, , \qquad |\det U|^2 =m^2 \, , 
\end{equation}
where $\det U$ is the determinant is of the complex  $2\times2$ matrix with entries $U_\alpha^I$.  Observe that  $|\det U|^2 = \det P = -P^2$, so  the condition $|\det U|^2 =m^2$ needed to solve the original mass-shell constraint is again a mass-shell constraint, but now expressed in terms of $U$. The choice of the upper sign in (\ref{Pus}) again corresponds to positive energy. 

We can now solve the fermionic constraint by writing $\lambda$ as 
\begin{equation}
\lambda_{\alpha\dot\alpha} = \frac{1}{m}U_\alpha^I U_{\dot\alpha}^J \psi_{IJ} + \frac{1}{m} P_{\alpha\dot\alpha} \xi\, , \qquad \left(\psi_{IJ}= \psi_{JI}\right)
\end{equation}
where the new anticommuting variables $\psi_{IJ}$ constitute an $SU(2)$  triplet.  As in the 3D case, the local supersymmetry transformation of $\lambda$ is now implied by that of $\xi$, so that $\psi_{IJ}$ is inert.  We may raise and lower $SU(2)$ indices with the invariant alternating tensor, which we do using the same conventions as for $Sl(2;\bC)$ spinor indices.  For example, 
\begin{equation}
\psi^I{}_J = \varepsilon^{IK} \psi_{KJ} = \psi^{IK}\varepsilon_{KJ}\, , 
\end{equation}
which  are the (anticommuting) entries of a traceless $2\times2$ Hermitian matrix. 

Substituting for $P$ and $\lambda$ in the action, we find the new Lagrangian
\begin{equation}
L= \dot U_\alpha^I W^\alpha_I + \dot{\bar U}_{\dot\alpha\, I} \bar W^{\dot\alpha\, I} + \frac{i}{4}\psi^I{}_J \dot\psi^J{}_I + \frac{d}{dt} \left(\cdots\right)\, , 
\end{equation}
where
\begin{eqnarray}\label{defsW}
W^\alpha_I &=& \mp\frac{1}{2} X^{\alpha\dot\alpha}\,  \bar U_{\dot\alpha\, I} \pm \frac{i}{2m^2} U^{\alpha J} \det \bar U \left(\psi_{IJ}\, \xi \mp \frac{1}{2}\psi_I{}^L\psi_{LJ}\right)\, ,  \nonumber \\
\bar W^{\dot\alpha\, I} &=& \mp \frac{1}{2} X^{\alpha\dot\alpha}\,  U_\alpha^I\ \mp\frac{i}{2m^2}\, \bar U^{\dot\alpha}_J \det U \left(\psi^{IJ}\xi \mp \frac{1}{2} \psi^I{}_L\psi^{LJ}\right)\, . 
\end{eqnarray}
Using these expressions, and the new mass-shell constraint
\begin{equation}
0= \varphi := |\det U|^2 -m^2 \, , 
\end{equation}
one may derive the identities
\begin{eqnarray}\label{Wids}
0 &\equiv&G :=  U_\alpha^I W^\alpha_I - \bar U_{\dot\alpha \, I}\bar W^{\dot\alpha \, I} \, , \nonumber\\
0 &\equiv& \Lambda^{IJ}  := U^{\alpha\, (I} W_\alpha^{J)} - \bar U^{\dot\alpha\, (I} \bar W_{\dot\alpha}^{ J)} +  \frac{i}{2}\psi^I{}_L\psi^{LJ} \, . 
\end{eqnarray}
Notice that $\Lambda^{IJ}=\Lambda^{JI}$ because of the anticommutativity of $\psi_{IJ}$. In order to promote the variables $W$ and $\bar W$ to the status of independent variables, these constraints must be imposed by Lagrange multipliers, along with the constraint $\varphi=0$. 

To simplify the final result, we first trade the anticommuting variables $\psi^I{}_J$ for a real anticommuting $3$-vector $\bfpsi$ by writing
\begin{equation}
\psi^I{}_J = \bfs^I{}_J \cdot \bfpsi\, . 
\end{equation}
The new action now takes the form 
\begin{equation}\label{newform}
S= \int dt\left\{ \dot U_\alpha^I W^\alpha_I + \dot{\bar U}_{\dot\alpha\, I} \bar W^{\dot\alpha\, I} + \frac{i}{2}\bfpsi\cdot \dot{\bfpsi} - \ell G -  s_{IJ}\Lambda^{IJ} - \rho \varphi\right\}\, , 
\end{equation}
where  $\ell$, $s_{IJ}= s_{JI}$ and $\rho$ are Lagrange multipliers for the constraints. Only the constraint functions $\Lambda^{IJ}$ involve the anticommuting 3-vector $\bfpsi$, and they now take the form
\begin{equation}\label{LambdaSig}
\Lambda^I{}_J = \Lambda_{(bos)}^I{}_J - i \bfs^I{}_J \cdot \bfS \, , \qquad \bfS= - \frac{i}{2} \bfpsi \times \bfpsi\, , 
\end{equation}
where $\Lambda_{(bos)}$ is the part independent of $\bfpsi$; it can be read off from (\ref{Wids}).  

The bosonic phase space is now parametrized by two pairs of complex 2-component spinors $Z^I = (U^I,W_I)$, and each pair is a complex 4-plet of $Sp(4;\bC)\cong U(2,2)$. 
Since $U(2,2)$ is (neglecting discrete factors) the product of $U(1)$ with the 4D conformal group, each of the $Z^I$ is a 4D twistor. The bosonic phase space is therefore parametrized 
by a pair of twistors, as for 3D but its real dimension is now $2\times 8=16$ and the 4D conformal invariance is broken by the mass-shell constraint. As there are a total of 
$5$ first-class constraints generating $5$ gauge invariances,  the bosonic dimension of the physical phase space is $16-10=6$. There are also $3$ real anticommuting coordinates 
not subject to any constraint or gauge invariance,  so the graded dimension  of the physical phase superspace is $(6|3)$, as it should be.

From the new action (\ref{newform}) we may read off the  canonical Poisson brackets. These are
\begin{equation}
\left\{ U_\alpha^I, W^\beta_J\right\}_{PB}= \delta_\alpha^\beta\delta^I_J\, , \qquad 
\left\{\bar U_{\dot\alpha\, I}, \bar W^{\dot\beta\, J} \right\}_{PB} = \delta_{\dot\alpha}^{\dot\beta}\delta_I^J\, , 
\end{equation}
and, for the components $\psi_i$ ($i=1,2,3$) of $\bfpsi$, 
\begin{equation}
\left\{\psi_i,\psi_j\right\}_{PB} = -i \delta_{ij}\, . 
\end{equation}
One can verify that the Poisson bracket algebra of the spin-shell constraint functions $(G,\Lambda^I{}_J)$ is $U(2)$. The anticommuting variables contribute only to the spin part of the $SU(2)$ generators, and one may easily check that their Poisson bracket algebra is
\begin{equation}
\left\{ \bfS_i ,\bfS_j\right\}_{PB} = \varepsilon_{ijk} \bfS_k \, .
\end{equation}

The constraint function $\varphi$ is manifestly $U(2)$ invariant, so it has zero Poisson brackets with the spin-shell constraints. The gauge invariance it generates has the following  transformations for parameter $b(t)$:
\begin{equation}
\delta_b W^\alpha_I =  b\,  U^\alpha_I \det \bar U\, , \quad \delta_b\bar W^{\dot\alpha\, I} = -b\,  \bar U^{\dot\alpha\, I} \det U\, , \qquad
\delta_b\rho = \dot b\, . 
\end{equation}
As for 3D, it is  important to take into account this gauge invariance (which is again equivalent to a time reparametrization)  when considering how $W$ transforms under the original local supersymmetry. The latter can  be deduced by using the transformations of $X$ and $\xi$ in the expressions of (\ref{defsW}):  this gives a $b$-transformation of the above type with $2m b= i\epsilon\xi$.  Thus the new twistor variables $(U,W,\psi)$ are gauge invariant with respect to all the original local symmetries modulo a gauge transformation generated by the new mass-shell constraint  function $\varphi$.  

Although we are calling  $G=0$ and $\Lambda^{IJ}=0$ the spin-shell constraints,  their relation to the particle's spin is not obvious because their inclusion in the action leads to the gauging of an apparently  {\it internal}  $U(2)$ symmetry.  In fact, the $U(1)$ constraint $G=0$ is not directly related to the particle's spin, but the $SU(2)$ constraint is, as becomes clear when one considers 
the  Pauli-Lubanski  spin vector.  In $Sl(2;\bC)$ spinor notation this is
\begin{equation}\label{PL}
{\cal W}_{\alpha\dot\alpha} = i\left({\cal J}_\alpha{}^\beta {\cal P}_{\beta\dot\alpha} -\bar{\cal J}_{\dot\alpha}{}^{\dot\beta}{\cal P}_{\alpha\dot\beta}\right)\, . 
\end{equation}
In twistor variables, the Lorentz Noether charges are 
\begin{eqnarray}
{\cal J}_\alpha{}^\beta &=& U_\alpha^I W^\beta_I - \frac{1}{2}\delta_\alpha^\beta\left(U_\gamma^KW^\gamma_K\right)\, , \nonumber \\
\bar{\cal J}_{\dot\alpha}{}^{\dot\beta} &=& \bar U_{\dot\alpha\, I} \bar W^{\dot\beta\, I} - 
\frac{1}{2}\delta_{\dot\alpha}^{\dot\beta} \left(\bar U_{\dot\gamma\, K}\bar W^{\dot\gamma\, K}\right)\, . 
\end{eqnarray}
Notice that there is no longer a contribution from  anticommuting variables, as expected from the fact that these are now Lorentz scalars. 
When these Poincar\'e Noether charges are substituted into the expression (\ref{PL}) one finds that 
\begin{equation}
{\cal W}_{\alpha\dot\alpha} = \pm\,  i \Lambda_{IJ}^{(bos)} U_\alpha^I \bar U_{\dot\alpha}^J\, , 
\end{equation}
where $\Lambda_{(bos)}$ is $\Lambda$ without the ``fermionic'' term. This shows that the bosonic particle has zero spin. 
For the spinning particle the additional spin term in $\Lambda_{IJ}$ is such that, when $\Lambda_{IJ}=0$, 
\begin{equation}
{\cal W}_{\alpha\dot\alpha} = \mp\,  U_\alpha^J\bar U_{\dot\alpha\, I}\,  \bfs^I{}_J\cdot \bfS \quad \Rightarrow \quad {\cal W}^2= m^2 \Sigma^2\, . 
\end{equation}
In the quantum theory,  $\Sigma^2$ equals $s(s -1)$ for an irreducible massive spin-$s$ representation of the Poincar\'e group, but to 
make use of this fact we must first quantize. 

To pass to the quantum theory we use Dirac's prescription to replace Poisson brackets of canonical variables by $-i$ times the (anti)commutator of their corresponding operators.  This yields the following  canonical anticommutation relation for the components of the operator $\hat{\bfpsi}$:
\begin{equation}
\left\{\hat\psi_i, \hat\psi_j\right\} = \delta_{ij} \quad \Rightarrow \quad \hat{\bfpsi} = \frac{1}{\sqrt{2}} \bft \quad \Rightarrow\quad \hat{\bfS}= \frac{1}{2} \bft \, , 
\end{equation}
where $\bft$ are Pauli matrices. It then follows that 
\begin{equation}
\hat\Sigma^2 = \frac{3}{4} \bI \quad \Rightarrow\quad s= \frac{1}{2}\, . 
\end{equation}
As expected, the quantum ${\cal N}=1$ spinning particle has spin $1/2$.

\subsection{Quantum theory for ${\cal N}\ge2$}

The twistor formulation  of the  ${\cal N}$-extended  spinning particle can be found by following  exactly the same procedure already explained for $D=3$. 
The resulting action is 
\begin{equation}\label{genN}
\!\!\!\!\! \!\!\!\!\! \!\!\!\!\! \!\!\!\!\! \!\!\!\!\! \!\!\!\!\! S = \int \! dt\left\{ \dot U_\alpha^I W^\alpha_I + \dot{\bar U}_{\dot\alpha\, I} \bar W^{\dot\alpha\, I} + \frac{i}{2}\bfpsi_a\cdot \dot{\bfpsi}_a 
- \, \ell G -  s_{ij}\Lambda^{IJ} - \rho\varphi  - f_{ab}\,  \bfpsi_a \cdot \bfpsi_b \right\}, 
\end{equation}
where $\Lambda^{IJ}$ is as given in (\ref{LambdaSig}) but now with 
\begin{equation}
\bfS = -\frac{i}{2}\sum_{a=1}^{\cal N}\bfpsi_a \times \bfpsi_a\, . 
\end{equation}
Proceeding as  we did for ${\cal N}=1$, we now find for ${\cal N}\ge2$ that 
\begin{equation}
\hat{\bfS} = \frac{1}{2}\left[ \bft \otimes \bI \otimes \bI \otimes \cdots  + \bI\otimes\bft \otimes\bI\otimes + \cdots  +
\bI\otimes\bI\otimes \cdots \otimes\bft\right]\, . 
\end{equation}
This acts on a reducible space of dimension $2^{\cal N}$. However, we still have to consider the $SO({\cal N})$ constraints; there are 
${\cal N}({\cal N}-1)$ of them. One implies that the state space is annihilated by the operator
\begin{equation}
\sum_{i=1}^3 \bft_i \otimes \bft_i \otimes\bI \cdots \otimes \bI\, . 
\end{equation}
However, this operator has eigenvalues $1$ and $-3$, so there is no state that satisfies the constraint. The theory is quantum inconsistent! 

This result is implicit in the conclusion  of   \cite{Howe:1989vn} that the massless spinning particle is inconsistent for ${\cal N}>2$ in odd spacetime dimensions\footnote{In even spacetime dimensions it describes a massless particle of spin $\frac{1}{2}{\cal N}$} because of a global anomaly (of the general type discussed in \cite{Elitzur:1985xj}). As pointed out in  \cite{Howe:1989vn}, the massive spinning particle in $D$ dimensions can be obtained by
a type of dimensional reduction from the massless spinning particle in $D+1$ dimensions, so we should expect the massive ${\cal N}>2$ spinning particle 
to be quantum inconsistent in {\it even} spacetime dimensions. This is what we find for $D=4$. 

As also pointed out in \cite{Howe:1989vn}, the global anomaly can be cancelled for $ {\cal N}=2$ by the WCS term.  We shall now recover this result from our twistor formulation of the model.  First, we introduce the complex anticommuting triplet
\begin{equation}
\bfpsi = \frac{1}{\sqrt{2}} \left(\bfpsi_1 + i\bfpsi_2\right)\, . 
\end{equation}
This leads to the ${\cal N}=2$ action 
\begin{eqnarray}
S&=& \int dt \Big\{ \dot U_\alpha^I W^\alpha_I + \dot{\bar U}_{\dot\alpha\, I} \bar W^{\dot\alpha\, I} + i\bar{\bfpsi} \cdot \dot{\bfpsi} - \ell G -  s_{ij}\Lambda^{IJ} - \rho\varphi  \nonumber \\
&&\qquad\qquad 
- \, \frac{1}{4} f \left[ \left(\bar{\bfpsi} \cdot \bfpsi  -  \bfpsi\cdot \bar{\bfpsi}\right) + 2 c\right]\Big\}\, , 
\end{eqnarray}
where $c$ is the coefficient of the WCS  term. Of course, $\bar{\bfpsi} \cdot \bfpsi  = -  \bfpsi\cdot \bar{\bfpsi}$  in the classical theory  but the
expression $(\bar{\bfpsi} \cdot \bfpsi  -  \bfpsi\cdot \bar{\bfpsi})$ yields the standard fermi oscillator operator ordering in the quantum theory. 

The canonical anticommutation relations of the operators $\hat{\bfpsi}$ and $\hat{\bfpsi}^\dagger$ are 
\begin{equation}
\left\{ \hat\psi_i, \hat\psi_j^\dagger\right\} = \delta_{ij}\, . 
\end{equation}
The $SO(2)$ constraint  therefore reduces to
\begin{equation}
n_1 +n_2 +n_3 - \frac{3}{2} + c =0\, , 
\end{equation}
where $n_i$ are the eigenvalues of the fermi number operators $\hat n_i\equiv {\hat\psi}_i^\dagger{\hat\psi}_i$ (no sum over $i$),  and 
we have allowed for the zero point contributions of each of the three fermi oscillators.  We now see that
\begin{equation}
c= \frac{3}{2} - k \, , \qquad k=0,1,2,3. 
\end{equation}
For each choice of $k$ we have 
\begin{equation}
\!\!\!\! \hat{\bfS}= -i \hat{\bfpsi} \times \hat{\bfpsi}{}^\dagger\quad \Rightarrow \quad \hat \Sigma^2 = (n_1-n_2)^2 + (n_2-n_3)^2+ (n_3-n_1)^2\, . 
\end{equation}
For $k=0,3$ (and hence $|c|=\frac{3}{2}$) all $n_i$ are equal, so there is a single polarization state with $\hat\Sigma^2=0$. These two cases describe a particle of zero spin.  For $k=1,2$ (and hence $|c|=\frac{1}{2}$), either one or two of the $n_i$ are zero,  and both cases give three polarization states with $\hat\Sigma^2=2$, which implies a particle of spin 1.  Since $c$ is non-zero in all  cases, we see that the WCS term is crucial  to quantum consistency.

\section{Massive 6D spinning particle}
\setcounter{equation}{0}

For the 6D spinning particle, we could use an $Sl(2;\bH)$ notation for spinors \cite{Kugo:1982bn} but it is simpler to use an $SU^*(4)$ notation\footnote{The $4\times4$ $SU^*(4)$ matrices are found from the $2\times2$ $Sl(2;\bH)$ matrices by using the $2\times2$ Pauli matrix representation of the algebra of quaternions.}.  In this notation, Lorentz 6-vectors are converted into antisymmetric  bi-spinors  by means of  a set of $6$ antisymmetric $4\times4$ matrices $\Sigma^m$ 
which can be chosen such that ($\alpha,\beta,\gamma,\delta=1,2,3,4$)
\begin{equation}\label{complete}
 \Sigma_m^{\alpha\beta}\Sigma^n_{\alpha\beta} = \delta_m^n \, , \qquad
\Sigma^m_{\alpha\beta} \Sigma_m^{\gamma\delta} = \delta_{[\alpha}^\gamma \delta_{\beta]}^\delta \, ,  
\end{equation}
where 
\begin{equation}\label{Sigup}
\Sigma_m^{\alpha\beta} := \frac{1}{2}  \varepsilon^{\alpha\beta\gamma\delta} \Sigma^n_{\gamma\delta} \eta_{mn} \, . 
\end{equation}
For this choice\footnote{Our choice of factors differs from those of \cite{Howe:1983fr}, where other aspects of the $SU^*(4)$ spinor notation are explained.} 
we have, for example, 
\begin{equation}\label{Pform}
\bP_{\alpha\beta} = \Sigma^m_{\alpha\beta} P_m\, , \qquad  P_m =  \Sigma_m^{\alpha\beta} \bP_{\alpha\beta}\, , 
\end{equation}
As suggested by the definition (\ref{Sigup}),  we may raise or lower antisymmetric pairs of spinor indices using the $SU^*(4)$ invariant alternating tensor.  For example, 
\begin{equation}
\bP^{\alpha\beta} := \frac{1}{2} \varepsilon^{\alpha\beta\gamma\delta} \bP_{\gamma\delta}\, . 
\end{equation}
We then find, in agreement with \cite{Mezincescu:2013nta,Routh:2015ifa}, that
 that 
\begin{equation}
\bP^{\alpha\beta}\bP_{\alpha\beta} =P^2\, . 
\end{equation}

In the above conventions,  the  action for the 6D massive spinning particle is 
\begin{equation}\label{spinning}
\!\!\!\!\!\!\!\!\!\!\!\!\!\!\!\!\!\!\!\!\!\! S= \int\! dt \left\{ \dot{\bX}^{\alpha\beta} \bP_{\alpha\beta} + \frac{i}{2}\lambda^{\alpha\beta} \dot \lambda_{\alpha\beta} + \frac{i}{2} \xi\dot\xi - \frac{1}{2} e\left(\bP^2+m^2\right) +i\zeta \left(\lambda^{\alpha\beta}\bP_{\alpha\beta} +m\xi\right)\right\}\, . 
\end{equation}
The  infinitesimal gauge transformations generated by the constraints are 
\begin{equation}
\delta_\epsilon X^{\alpha\beta} = -i\epsilon\lambda^{\alpha\beta}\, , \qquad \delta_\epsilon\lambda_{\alpha\beta} =\epsilon\, \bP_{\alpha\beta}\, \qquad \delta_\epsilon \xi= m\epsilon\, . 
\end{equation}
The Lorentz Noether charges are 
\begin{equation}
{\cal J}_{\alpha}{}^{\beta} = 2 \bP_{\alpha\gamma}\bX^{\beta\gamma} - \frac12 \delta_{\alpha}^{\beta}\, \bP_{\gamma\delta} \bX^{\gamma\delta} 
+ i \lambda_{\alpha\gamma}\lambda^{\beta \gamma} \,. 
\end{equation}

To pass to the twistor form of the action, we first solve the mass-shell constraint as in \cite{Routh:2015ifa} by  
setting\footnote{The sign of the energy now depends on the choice of the $USp(4)$-invariant tensor $\Omega$. Also,  notice the sign of $\det \bU$ on the mass shell. }
\begin{equation}\label{Psq1}
\bP_{\alpha\beta} = \frac{1}{2} \bU_\alpha^I \bU_\beta^J\, \Omega_{JI} \, ,\qquad \det \bU=-m^2 \, , 
\end{equation}
where $\bU$ is a 4-plet $(I=1,2,3,4)$ of  $SU^*(4)$ spinors, and $\Omega$ is the antisymmetric $USp(4)$-invariant matrix (normalized such that $\det\Omega=1$). We use $\Omega$ to raise and lower indices using the same conventions that we used previously for $Sl(2;\bR)$ and $Sl(2;\bC)$. To verify that the mass-shell constraint is solved, one needs the identity $3\Omega_{I[J} \Omega_{KL]} \equiv  \epsilon_{IJKL}$,  where $\epsilon_{IJKL}$ is the  invariant alternating tensor of $USp(4)$. The invariant tensor $\epsilon^{IJKL}$ is then defined by raising 
indices, which implies that  $\epsilon^{1234}=1$ given (as we assume) that $\epsilon_{1234}=1$.
As for the 3D and 4D cases, the constraint on the determinant of $U$ can be viewed as a new mass-shell constraint. 

Before proceeding it is convenient to define
\begin{equation}
\bV_I^{\alpha} = \frac{1}{6m} \epsilon_{IJKL}\epsilon^{\alpha\beta\gamma\delta}\bU^J_{\beta}\bU^K_{\gamma}\bU^L_{\delta} \, . 
\end{equation}
This is a new $USp(4)$ $4$-plet of commuting $SU^*(4)$ spinors of opposite chirality to $\bU$, and the two are inverses of each other, up to factors, since
\begin{equation}
\bV^\alpha_I \bU_\alpha^J  = -m\delta_I^J\, , \qquad \bV^\alpha_I \bU_\beta^I = -m\delta^\alpha_\beta\, \qquad (\det \bU = -m^2)\, . 
\end{equation}
It can be shown, again on the  surface  $\det \bU=-m^2$,  that  \cite{Routh:2015ifa}
\begin{equation}
\bP^{\alpha\beta} = -\frac12\, \bV_I^{\alpha}\bV_J^{\beta}\,  \Omega^{JI} \, ,  \qquad (\det \bU = -m^2)\, . 
\end{equation}

Next, we solve the fermionic constraint by setting
\begin{equation}\label{solvef}
\lambda_{\alpha\beta} = \frac{1}{\sqrt{2}\, m} \bU_{\alpha}^I \bU_{\beta}^J \, \psi_{IJ} + \frac{1}{m} \bP_{\alpha\beta} \xi\, , 
\end{equation}
where $\psi_{IJ}$ is antisymmetric and $\Omega$-tracefree, and hence has five independent components. As for the 3D and 4D cases, the supersymmetry transformation of $\lambda$ is now implied by that of $\xi$, so that the new anticommuting variables $\psi$ are inert.  A useful alternative, but equivalent, expression for $\lambda$ is 
\begin{equation}
\lambda^{\alpha\beta}= \frac{1}{\sqrt{2}\, m} \bV^\alpha_I\bV^\beta_J \psi^{IJ} + \frac{1}{m}\bP^{\alpha\beta} \xi\, . 
\end{equation}
To prove equivalence of this expression to (\ref{solvef}) one needs the relation 
\begin{equation}
\psi^{IJ}= - \frac{1}{2}\epsilon^{IJKL}\psi_{KL}\, .  
\end{equation}
The left hand side  is defined by raising indices with $\Omega$. To show that this equals the right hand side one uses the identity
$\epsilon^{IJKL} = \Omega^{I[J}\Omega^{KL]}$ and the fact that $\psi_{IJ}$ is both antisymmetric and 
$\Omega$-traceless\footnote{It is not consistent to use the alternating invariant tensor of  $USp(4)$ to raise or lower antisymmetric pairs of  $USp(4)$ spinor  indices because a different sign would then be needed to apply this to $\Omega$ itself, as  follows from identity $\Omega^{IJ} \equiv \frac{1}{2}\epsilon^{IJKL}\Omega_{KL}$.}.

Substituting for $\bP$ and $\lambda$, one  finds that
\begin{equation}\label{halfway}
\dot{\bX}^{\alpha\beta} \bP_{\alpha\beta} + \frac{i}{2} \lambda_{\alpha\beta}\dot\lambda^{\alpha\beta} + \frac{i}{2} \xi \dot \xi  = 
\dot{\bU}_\alpha^I\bW^\alpha_I + \frac{i}{4} \psi^{IJ} \dot\psi_{IJ} 
+ \frac{d}{dt} \left(\dots\right)\, , 
\end{equation}
where\footnote{We choose the overall sign of $\bW$ to be opposite to that chosen in \cite{Routh:2015ifa} so that the form of the action is similar to the 3D and 4D cases deduced in previous sections.}
\begin{equation}\label{newW}
W^\alpha_I =-\bX^{\alpha\beta}U^J_{\beta}\Omega_{JI}  - \frac{i}{\sqrt{2}\, m} \bV^\alpha_K \psi^K{}_I \xi - \frac{i}{2m} \bV^\alpha_K  \, \psi^{KJ}\psi_{IJ}\, . 
\end{equation}
This expression  implies the identity
\begin{equation}
0\equiv  \bU^{(I}_{\alpha}  \bW^{\alpha J)} - \frac{i}{2} \psi^{KI} \psi_K{}^{J} \equiv \Lambda^{IJ} \, , 
\end{equation}
which becomes a constraint imposed by a Lagrange multiplier in the twistor form of the action in which $\bW$ is an independent variable. This action is 
\begin{equation}\label{spinaction}
S=  \int dt \, \left\{ \dot{\bU}^I_{\alpha}  \bW_I^{\alpha} + \frac{i}{4}\psi_{IJ}\dot \psi^{IJ}  - s_{IJ} \Lambda^{IJ} - \rho\, \varphi\right\}\, . 
\end{equation}
As anticipated, the only surviving anticommuting phase space variables  are the five independent components of  $\psi_{IJ}$. 

The bosonic phase space is now parametrized by the spinor pair $(U^I,W_I)$, with each spinor in the $({\bf 4},{\bf 4})$ representation of $SU^*(4)\times USp(4)$; equivalently, 
each spinor is a pair of $Sl(2;\bH)$ spinors, and the spinor pair $(U^I,W_I)$ is equivalent to a pair of  $4$-component quaternionic spinors in the $({\bf 4},{\bf 2})$ representation of 
$Sp(4;\bH) \times U(2;\bH)$. A single $4$-component quaternionic $4$-plet of $Sp(4;\bH)$ is a spinor of the 6D conformal group, 
and hence a 6D twistor, so the bosonic phase space of the massive 6D particle described by the action (\ref{spinaction}) is parametrized by a pair of twistors, exactly as we found earlier for 
$D=3,4$.  The real dimension of this space is now $4\times 8 =32$ but these variables are subject to $10+1=11$ first class constraints, which generate $11$ gauge invariances, so the physical bosonic dimension of phase space is $32-2\times 11 =10$. There are also $5$ real anticommuting variables, not subject to any constraints, so the graded real dimension of the physical phase superspace is 
$(10|5)$, as expected from our starting point.

The new mass-shell constraint $\varphi=0$ is associated with the following gauge invariance with parameter $b(t)$:
\begin{equation}
\delta_b \bW^I_\alpha = - mb V^\alpha_I \, , \qquad \delta_b \rho = \dot b\, . 
\end{equation}
As shown in \cite{Routh:2015ifa}, this is equivalent to a time reparametrization, and as in the 3D and 4D cases, a $b$-gauge transformation 
of $\bW$ is induced by a local worldline supersymmetry transformation of $\bX$ and $\xi$ in the expression (\ref{newW}). Specifically, one finds that 
\begin{equation}
\delta_\epsilon \bW^\alpha_I= i\epsilon\xi V^\alpha_I\, \, . 
\end{equation}
The twistor variables are therefore gauge-invariant under the original gauge transformations modulo a $b$-gauge transformation with parameter $b= -i\epsilon\xi/m$. 

We may  simplify the action (\ref{spinaction})  by writing 
\begin{equation}
\psi_{IJ} = \frac{1}{2} \left(\gamma^a\right)_{IJ} \psi_a \, ,
\end{equation}
where  $\psi_a$ is an anti-commuting 5-vector, and $(\gamma^a)_I{}^J$ ($a=1,\dots,5$) are the five $4\times4$ ${\rm Spin}(5)$ Dirac matrices satisfying 
\begin{equation}
\left\{\gamma^a,\gamma^b\right\} = 2 \delta^{ab}\, . 
\end{equation}
For the choice $\Omega= \bI_2\otimes i\sigma_2$, a basis for these matrices is 
\begin{equation}
\gamma^a =\left\{ \sigma_2\otimes \sigma_1\, ,  \sigma_2\otimes\sigma_2\, , \sigma_2\otimes \sigma_3\, , 
 \sigma_1\otimes\bI_2\, ,  \sigma_3\otimes\bI_2\right\}\, ,  
\end{equation}
and the five antisymmetric matrices with entries $\gamma^a_{IJ}$ are  $\gamma^a\Omega$. The action becomes
\begin{equation}\label{equivaction}
S=  \int dt \, \left\{ \dot{\bU}^I_{\alpha} \bW_I^{\alpha} + \frac{i}{4}\psi_a\dot \psi_a  - s_{IJ} \Lambda^{IJ} - \rho\, \varphi\right\}\, , 
\end{equation}
where now
\begin{equation}
\Lambda^{IJ} =  \bU^{(I}_{\alpha} \Omega^{J)K} \bW^{\alpha}_K + \frac{i}{8} \left(\gamma^{ab}\right)^{IJ} \psi_a\psi_b \qquad \left(\gamma^{ab} = \gamma^{[a}\gamma^{b]}\right)\, . 
\end{equation}
The Lorentz  Noether charge in the twistor variables are
\begin{equation}
{\cal J}_{\alpha}{}^{\beta} = \bU^I_{\alpha}\bW^{\beta}_I  - \frac{1}{4} \delta_\alpha^\beta \left(\bU_\gamma^IW^\gamma_I\right)\,. 
\end{equation}
As expected, there  is no fermion bilinear term  because the anticommuting variables are Lorentz scalars. 

In the quantum theory, the  Poisson bracket relations of the anticommuting variables $\psi_a$ become the canonical anticommutation relations
\begin{equation}
\left\{\hat\psi_a,\hat\psi_b\right\} = 2\delta_{ab} \quad \Rightarrow \quad  \hat \psi_a=  \Gamma_a\, , 
\end{equation}
where $\Gamma_a$ are another set of  ${\rm Spin}(5)$ Dirac matrices. This is exactly what one finds in the non-relativistic limit for a particle
of spin $\frac{1}{2}$ in 5-dimensional Euclidean space, so we confirm that the quantum  ${\cal N}=1$ 6D massive spinning particle has spin $\frac{1}{2}$.

The twistor form of the  6D massive spinning particle action for ${\cal N}>1$ can be obtained exactly  in the way described earlier for $D=4$.  The ${\cal N}=2$ case is the one of most interest because  the ${\cal N}>2$ cases are inconsistent as quantum theories. We pass over the details since the the end  results are known from the work of  \cite{Howe:1989vn}. The
procedure is similar to that already described for 4D but the description of the results obtained  involves consideration of  the 6D Pauli-Lubanski tensors given in \cite{Routh:2015ifa}, which 
goes beyond the scope of this paper. 

\section{Discussion}
\setcounter{equation}{0}

In this paper we have shown how the massive ``spinning particle'' (with local worldline supersymmetry)  may be reformulated in twistor variables for spacetime dimension $D=3,4,6$. 
Our results duplicate those of  \cite{Mezincescu:2013nta} for $D=3$ but our new construction generalises directly to $D=4,6$.

A feature  of the twistorial action is that the anticommuting spin variables appear exactly as they do in the non-relativistic limit!  
This is possible because {\it the twistor variables are invariant under the local supersymmetry  of the original action}.  This is also true for the massless spinning particle but the results are more striking for non-zero mass, partly because it  is only in this case that one can consider a non-relativistic limit, and partly because the spin-shell group (which coincides, or ``almost coincides'' with Wigner's little group)  is larger  for massive particles. 

It is also true that the original variables are invariant under the gauge transformations that act on the twistor variables. In this sense, the two formulations are dual, sharing physical content but differing in the extra variables used to ensure manifest Lorentz invariance.  Actually, for massive particles  both formulations share a common gauge invariance, generated by a Hamiltonian constraint, equivalent in both formulations  to a time reparametrization. The mass-shell constraint in one formulation is simply exchanged for a mass-shell constraint in the other.  Invariance of one set of variables with respect to the gauge invariances of the other must therefore be understood to be ``modulo'' a time-reparametrization invariance.

Because of the simple non-relativistic nature of the ``fermionic''  terms in the twistorial version of the massive spinning particle action, the analysis of the 
implications for the quantum theory is simplified. We illustrated this fact by an analysis of the 4D ${\cal N}$-extended spinning particle. The results
are either known or implicit in earlier work, but we were able to simply confirm both that the ${\cal N}=2$ quantum spinning particle  has either spin zero or spin one, and that  the ${\cal N}>2$ massive spinning particle is quantum inconsistent in even spacetime dimensions.

Implicit in our results for the massive spinning particle is a twistor description of the {\it massless} spinning particle for $D=3,4,6$, obtained by setting the mass to zero. One would expect to be able to simplify the action in this case so as to parametrize the bosonic phase space by the components of a single twistor, and we have spelt out the details of the procedure that achieves this for $D=3$. In all cases, 
the graded dimension of the physical phase space is the same as that found by taking the massless limit of the massive spinning particle in its standard phase space formulation, but this limit
does not yield the usual massless spinning particle action: there remains an additional  anticommuting ``spectator'' variable. This discontinuity in the massless limit of the spinning particle
appears not to have been commented on previously. 

In our twistor construction, we allowed for either sign of the energy when solving the mass-shell constraint on the $D$-momentum. A notable feature of our results for the twistorial action is that  this sign choice does not appear in it.  This does not happen for the superparticle, where the sign of the fermion kinetic terms is correlated with the sign of the energy, as pointed out  in \cite{Gauntlett:1990xq} in the context of a comparison of gauge-fixed superparticle and spinning particle actions. In an appendix we have confirmed this correlation for the massive 4D superparticle from a supertwistor form of its action constructed along the lines of the 6D case in \cite{Routh:2015ifa}.

Our results for  massive spinning particles  fit nicely with the association of the spacetime dimensions $D=3,4,6$ with the division algebras 
$\bR,\bC,\bH$, and this suggests a possible octonionic extension to $D=10$.  We expect the massive spinning particle to have a twistor formulation for which 
the  bosonic variables are the components of  a pair of octonionic twistors, i.e.  two 4-plets of  $Sp(4;\bO)$. Appropriately defined \cite{Chung:1987in}, this is the
$D=10$ conformal group. We also expect  22 spin-shell constraints in the form of an  anti-hermitian $2\times 2$ matrix over $\bO$, and  (from a reading of  \cite{Chung:1987in})
we would expect this to imply a $U(2;\bO)\cong {\rm Spin}(9)$ gauge invariance (since $22+14=36$). If this is correct, and taking into account a mass-shell constraint, we would have 
a physical phase superspace of bosonic dimension $64 - 2(22 +1)=18$. Given that  the fermionic variables are again the entries of  a  traceless hermitian $2\times 2$ matrix,  now over $\bO$
and presumably equivalent to a ${\bf 9}$ of  ${\rm Spin}(9)$,  we would then have  a physical phase superspace of graded dimension $(18|9)$, as required.

\section{Appendix: Massive 4D superparticle}
\setcounter{equation}{0}
\renewcommand{\theequation}{A-\arabic{equation}}

In this appendix we present the twistor form of the minimal massive 4D superparticle  in the notation of  this paper,  following the twistor construction of  the 6D superparticle in  \cite{Routh:2015ifa}.  The action is 
\begin{equation}\label{Spartact}
S= \int\! dt \left\{-\frac{1}{2} \left(\dot X^{\alpha\dot\alpha} + 
i\bar\theta^{\dot\alpha} \dot\theta^\alpha -i \dot{\bar\theta}^{\dot\alpha} \theta^\alpha\right)P_{\alpha\dot\alpha} 
- \frac{1}{2} e\left(P^2 +m^2 \right) \right\}\, .
\end{equation}
Since the phase superspace has graded dimension $(8|4)$ and there is one (first-class) constraint generating a gauge invariance, the physical phase superspace has 
graded dimension $(6|4)$. 

The action is manifestly  invariant under the spacetime supersymmetry transformations
\begin{equation}
\delta X^{\alpha\dot\alpha} =- i \bar\epsilon^{\dot\alpha}\theta^\alpha - i\epsilon^\alpha \bar\theta^{\dot\alpha}\, ,
 \qquad \delta\theta^\alpha= \epsilon^\alpha\, . 
\end{equation}
The  anticommuting Noether  charges are
\begin{equation}
Q_\alpha = P_{\alpha\dot\alpha}\bar\theta^{\dot\alpha}\, , \qquad \bar Q_{\dot\alpha} = P_{\alpha\dot\alpha} \theta^\alpha\, . 
\end{equation}
There are also hidden supersymmetry charges \cite{Mezincescu:2014zba}:
\begin{equation}
\tilde Q^\alpha = m\theta^\alpha\, , \qquad \bar{\tilde Q}^{\dot\alpha} = m \bar\theta^{\dot\alpha}\, . 
\end{equation}

The Poisson brackets of canonical variables follow directly from the action. The only non-zero one that we will need here is
\begin{equation}
\left\{\theta^\alpha,\bar\theta^{\dot\alpha}\right\}_{PB}  =-i P^{\alpha\dot\alpha}/P^2\, . 
\end{equation}
Using this, the full algebra of  supersymmetry charges, manifest and hidden,  is found to be the  BPS-saturated ${\cal N}=2$ supersymmetry algebra 
\begin{eqnarray}
\left\{Q_\alpha,\bar Q_{\dot\alpha}\right\}_{PB} &=& iP_{\alpha\dot\alpha} \, , \qquad 
\left\{\tilde Q^\alpha,\bar {\tilde Q}^{\dot\alpha}\right\}_{PB} = iP^{\alpha\dot\alpha}\, , \nonumber \\
\left\{Q_\alpha, \tilde Q^\beta \right\}_{PB} &=&  im\,  \delta_\alpha^\beta\, , \qquad 
\left\{ \bar Q_{\dot\alpha}, \bar{\tilde Q}^{\dot \beta} \right\}_{PB} = im\,  \delta_{\dot\alpha}^{\dot\beta}\, . 
\end{eqnarray}

To pass to the supertwistor form of the action, we solve the mass-shell constraint as we did for the 4D spinning particle.
Substitution for $P$ then yields the new Lagrangian 
\begin{equation}
L= \left[ \dot U_\alpha^I W^\alpha_I + \dot {\bar U}_{\dot\alpha\, I} \bar W^{\dot\alpha\, I} \pm i\bar\mu_I\dot\mu^I \right]
\end{equation}
where 
\begin{equation}
\mu^I = \theta^\alpha U_\alpha^I\, , \qquad \bar\mu_I = \bar\theta^{\dot\alpha} \bar U_{\dot\alpha\, I}\, , 
\end{equation}
and 
\begin{equation}
W^\alpha_I= \mp \frac{1}{2} X^{\dot\alpha\alpha}\bar U_{\dot\alpha\, I} \mp \frac{i}{2} \bar\mu_I \theta^\alpha\, , \qquad 
\bar W^{\dot\alpha\, I} = \mp\frac{1}{2} X^{\dot\alpha\alpha}U_\alpha^I \mp \frac{i}{2}\mu^I\bar\theta^{\dot\alpha}\, . 
\end{equation}
From these expressions we deduce the identity
\begin{equation}
0 \equiv G^I{}_J := U_\alpha^I W^\alpha_J - \bar U_{\dot\alpha\, J} \bar W^{\dot\alpha\, I} \mp  i \mu^I\bar\mu_J \, . 
\end{equation}
This becomes a $U(2)$ constraint when we promote $W$ to an independent variable. This leads to the new action
\begin{equation}
S=   \int\! dt \left\{ U_\alpha^I \dot W^\alpha_I + \bar U_{\dot\alpha\, I} \bar W^{\dot\alpha\, I} \pm  i\bar\mu_I\dot\mu^I  - s_J{}^I G^I{}_J - \rho \varphi\right\}\, , 
\end{equation}
where $\varphi = |\det U|^2 -m^2$, as for the massive spinning particle. 

As a check on this action, one may verify that the  (graded) dimension of the  physical phase superspace is still  $(6|4)$. The $SU(2)$ doublet of twistors $(U,W)$ have $2\times4=8$ complex components, giving a real bosonic dimension of 16, but there are $4$ spin-shell constraints, with spin-shell algebra $U(2)$, and $1$ further mass-shell-type constraint; all are first class  so the physical bosonic dimension of $16-2\times 5 =6$. The new anticommuting variable $\mu^I$ is a complex doublet of $U(2)$, and there are no fermionic constraints, so the physical fermionic dimension is $4$. 

A crucial feature of this superparticle action, which is required by spacetime supersymmetry \cite{Gauntlett:1990xq}, is the correlation between the sign of the energy and the sign of the ``fermionic kinetic term''. 

\subsection*{Acknowledgments}

We are grateful to Paul Howe for helpful correspondence,  and to Alex Arvanitakis and Joaquim Gomis for discussions. LM acknowledges partial support from the  National Science Foundation Award PHY-1214521. PKT acknowledges support from the UK Science and Technology Facilities Council (grant ST/L000385/1). AJR is supported by a grant from the London Mathematical Society, and he thanks the University of Groningen for hospitality during the writing of this paper. LM and PKT are grateful for the hospitality of  the Pedro Pascual Benasque Center for Science, where part of this work was done.

\bigskip


\providecommand{\href}[2]{#2}\begingroup\raggedright\endgroup

\end{document}